\documentclass[prd,twocolumn,nopacs,floatfix,amsmath,nofootinbib,amssymb,floatfix]{revtex4}
\usepackage{graphicx,color,dcolumn,booktabs,bm}
\usepackage{longtable}
\usepackage{pdflscape}
\usepackage{pdfpages}
\usepackage{txfonts}
\usepackage{overpic}
\usepackage{amssymb}
\usepackage{makecell}
\usepackage{indentfirst}
\usepackage{feynmf}   
\usepackage{slashed}  
\usepackage{cases}
\usepackage{color}
\usepackage{multirow}
\usepackage{threeparttable}
\usepackage{epstopdf}
\usepackage{enumerate}
\usepackage{ulem}
\usepackage[figuresright]{rotating}
\usepackage[colorlinks,
            citecolor=blue,
            anchorcolor=black,
            menucolor=black,
            linkcolor=black,
            filecolor=black,
            runcolor=black,
            urlcolor=blue,
            frenchlinks=true]{hyperref}

\graphicspath{{Figures/}{../图片/}} %

\allowdisplaybreaks

\begin{document}
\title{$K_1^*(1870)$ and $3S$-$2D$ mixing in the strange meson sector}
\author{Ting-Yan Li$^{1,2,3,4}$}\email{lity2023@lzu.edu.cn}
\author{Zi-Yue Bai$^{1,2,3,4}$}\email{baiziyue@lzu.edu.cn}
\author{Xiang Liu$^{1,2,3,4}$}\email{xiangliu@lzu.edu.cn}
\affiliation{
$^1$School of Physical Science and Technology, Lanzhou University, Lanzhou 730000, China\\
$^2$Lanzhou Center for Theoretical Physics,
Key Laboratory of Theoretical Physics of Gansu Province,
Key Laboratory of Quantum Theory and Applications of MoE,
Gansu Provincial Research Center for Basic Disciplines of Quantum Physics, Lanzhou University, Lanzhou 730000, China\\
$^3$MoE Frontiers Science Center for Rare Isotopes, Lanzhou University, Lanzhou 730000, China\\
$^4$Research Center for Hadron and CSR Physics, Lanzhou University and Institute of Modern Physics of CAS, Lanzhou 730000, China}

\date{\today}
\begin{abstract}
Very recently, the BESIII Collaboration reported the first observation
of a $J^P=1^-$ strange resonance in the $K\eta'$ system, the
$K_1^*(1870)$. We examine whether this state can be accommodated in the
kaon meson family. Although it cannot be assigned as a pure $3^3S_1$ or
$2^3D_1$ state, we show that it can be described as the lower state of
$3S$-$2D$ mixing, whose mass agrees with the measured value, and we
predict its higher-mass partner. Using the quark pair creation model,
we calculate the OZI-allowed decay properties of both \textcolor{black}{mixed} states;
the total decay width of the lower state is consistent with the
measured width of the $K_1^*(1870)$. We note that the subordinate
$K\eta'$ decay mode of this state appears to contradict its observed
$K\eta'$ decay channel, which calls for further study. We further
investigate the production of these two states via the $Kp$ reaction,
which may provide valuable information for future experimental
searches.
\end{abstract}

\maketitle

\section{INTRODUCTION} \label{section1}

Very recently, the BESIII Collaboration reported the first observation
of a strange resonance with $J^P=1^-$ in the $K\eta'$ system, denoted
as the $K_1^*(1870)$ \cite{BESIII:2026K1star1870}. This result comes from
a partial-wave analysis of $\psi(2S)\to\gamma\chi_{c1}$,
$\chi_{c1}\to K^+K^-\eta'$, using $\eta'\to\pi^+\pi^-\gamma$ and
$\eta'\to\pi^+\pi^-\eta$. With a sample of
$(2712.4\pm14.3)\times10^6$ $\psi(2S)$ events, the $J^P=1^-$ amplitude
is favored at $11.0\sigma$, while the tested $0^+$ and $2^+$ hypotheses
have significances of $3.2\sigma$ and $3.7\sigma$, respectively. The
measured mass and width are $M=(1869.0\pm14.6\pm18.6)$ MeV and
$\Gamma=(167.3\pm34.0\pm24.2)$ MeV, and the branching fraction for
$\chi_{c1}\to K^\mp K_1^*(1870)^\pm\to K^+K^-\eta'$ is
$(1.48\pm0.30\pm0.33)\times10^{-5}$
\cite{BESIII:2026K1star1870}. This observation provides useful input for
constructing the kaon meson family or identifying hybrid states.

If we assign the reported $K_1^*(1870)$ as a strange hybrid, we notice
that the predicted mass of the corresponding vector strange hybrid is
1720 MeV, which is lower than that of the $K_1^*(1870)$, although its
width is close to that of the $K_1^*(1870)$. In fact, we find that the
$K\eta'$ decay of the vector strange hybrid can be ignored, as presented
in Refs.~\cite{Chen:2023hybrid,Chen:2026hybriderratum}, which poses a challenge to categorizing the
$K_1^*(1870)$ as a hybrid.

Of course, we may test the possibility of the $K_1^*(1870)$ within the
strange meson assignment. As shown in Table~\ref{tab:k1star_assignments},
the observed $K_1^*(1870)$ \textcolor{black}{lies below the predicted masses of} the $3^3S_1$ and $2^3D_1$
states, according to the theoretical results in
Refs.~\cite{Pang:2017dlw,Feng:2021qmt}. Although it is difficult to
explain the $K_1^*(1870)$ as the $3^3S_1$ or $2^3D_1$ kaon state, there
exists a possibility of placing the $K_1^*(1870)$ into the kaon meson
family if the $3S$-$2D$ mixing is considered.

$S$-$D$ mixing is \textcolor{black}{widely studied} in meson families. As proposed in
Ref.~\cite{Rosner:2001mix}, the $2S$-$1D$ mixing scheme is applied to clarify the
puzzling decay properties of the charmonia $\psi(3686)$ and
$\psi(3770)$. \textcolor{black}{To interpret higher vector charmonium and bottomonium states}, $4S$-$3D$ mixing is proposed
in Refs.~\cite{Wang:2019mix,Wang:2023psi4500,Wang:2024psiY,
Peng:2024highercharmonia,Li:2021mix,Bai:2022upsilon,
Li:2022upsilontrans,Liu:2024xb,Liu:2024bottomtrans}, which may quantitatively originate from the
coupled-channel effect. There are a series of studies of $S$-$D$ mixing
for higher charmonia \cite{Barnes:2005charm,Li:2009screened,Duan:2021MGI,Wang:2019mix}. This mixing scheme can also occur in
\textcolor{black}{bottomonium} and vector light-flavor mesons \cite{Li:2021mix,Piotrowska:2017}. Interested readers can
consult the recent review article \cite{Bai:2026atm}.

These experiences tell us that it is a good opportunity to investigate
$3S$-$2D$ mixing of the kaon meson; \textcolor{black}{here we explore its application
to the newly observed $K_1^*(1870)$}.

Our study shows that the lower state of $3S$-$2D$ mixing of the kaon
meson can match the mass of the $K_1^*(1870)$. We predict its partner
with a higher mass. This information is not enough to identify the
properties of the $K_1^*(1870)$. We then present the
Okubo--Zweig--Iizuka (OZI)-allowed decay properties of these two states
through the quark pair creation (QPC) model \cite{Pang:2017dlw}. The calculated total decay
width of the lower state of $3S$-$2D$ mixing of the kaon meson is
consistent with the measured width of the $K_1^*(1870)$. We note that
the $3S$-$2D$ mixing state corresponding to the $K_1^*(1870)$ has a
subordinate decay mode $K\eta'$, which seems to contradict the fact that
the observed decay channel of the $K_1^*(1870)$ is $K\eta'$. This should
be clarified by further studies.

\begin{table}[htbp]
\caption{BESIII parameters for \(K_1^*(1870)\) compared with representative
pure-state predictions. Masses and total widths are in MeV; the two BESIII
uncertainties are statistical and systematic, respectively.}
\label{tab:k1star_assignments}
\centering
\footnotesize
\setlength{\tabcolsep}{3pt}
\begin{threeparttable}
\begin{ruledtabular}
\begin{tabular}{lcc}
Assignment & \(M\) & \(\Gamma\) \\
\hline
$K_1^*(1870)$ \cite{BESIII:2026K1star1870} &

$1869.0\pm14.6\pm18.6$ &

$167.3\pm34.0\pm24.2$\\
\(K(3\,{}^3S_1)\) \cite{Pang:2017dlw} & 1983 & 393 \\
\(K(2\,{}^3D_1)\) \cite{Pang:2017dlw} & 2127 & 537\tnote{a}\\
\(K(3\,{}^3S_1)\) \cite{Feng:2021qmt} & 1950 & 176 \\
\(1^-\) strange hybrid \cite{Chen:2023hybrid,Chen:2026hybriderratum} & 1720 & 180 \\
\end{tabular}
\end{ruledtabular}
\begin{tablenotes}[flushleft]
\footnotesize
\item[a] The $K(2\,{}^3D_1)$ width of 537 MeV is calculated in the
present work using the model parameters of Ref.~\cite{Pang:2017dlw};
this width was not reported in that reference.
\end{tablenotes}
\end{threeparttable}
\end{table}

In the last century, the $Kp$ reaction was an effective approach to
search for kaon mesons (see Particle Data Group (PDG) \cite{ParticleDataGroup:2024cfk}).
In the past years, the COMPASS Collaboration has played an important
role in searching for mesonic states. In particular, COMPASS recently
observed a signal of the strange $K(1690)$ state in
$K^-+p\to K^-\pi^-\pi^+ +p$ \cite{COMPASS:2025wkw}, which inspired discussion of the
properties of the observed $K(1690)$ {\hypersetup{citecolor=black}\cite{Chen:2026k1690}}. Thus, the $Kp$ reaction
can be applied to investigate the production of the $K_1^*(1870)$,
where the obtained partial decay information can serve as input for the
concrete calculation of the production via the $Kp$ reaction. We focus
not only on the production of the $K_1^*(1870)$ but also on the
production of its partner through the $Kp$ reaction. This may provide
valuable information for further experimental searches for these
strange states and can serve as a complement to the study of
$K_1^*(1870)$ production in $\chi_{c1}$ decay.

This paper is organized as follows. In Sec.~\ref{section2}, we
introduce the $3S$-$2D$ mixing scheme for the kaon family and
determine the mixing angle from the measured mass of the
$K_1^*(1870)$, together with the predicted mass of its orthogonal
partner. We also present the OZI-allowed decay
properties of the two mixed states calculated within the QPC model,
including their partial widths and the $K\eta'$ branching fractions.
In Sec.~\ref{section3}, we calculate the pion-exchange
production of these states in $K^-p$ reactions within an effective
Lagrangian approach with a Reggeized propagator, and give the total
cross sections \textcolor{black}{and} angular distributions. Section~\ref{section4} is devoted to a summary and
discussion.

\section{\texorpdfstring{THE $3S$-$2D$ MIXING SCHEME FOR THE $K_1^*(1870)$}{THE 3S-2D MIXING SCHEME FOR THE K1*(1870)}}
\label{section2}

We employ the modified Godfrey--Isgur (MGI) model, which is based on
the relativized quark model of Ref.~\cite{Godfrey:1985xj}. Its
Hamiltonian contains the relativistic kinetic energies of the
constituent quark and antiquark, together with the central, hyperfine,
and spin-orbit interactions. The linear confining term $br+c$ is
replaced by the screened form $b(1-e^{-\mu r})/\mu+c$, where $\mu$ is
the screening parameter. This screened-potential description was
developed in quarkonium studies \cite{Ding:1993screened,Li:2009screened}
and implemented in relativized quark models for light and heavy mesons
\cite{Song:2015nia,Song:2015fha,Pang:2017dlw,Li:2022khh,
Pang:2019prediction,Duan:2021MGI}.

 The MGI parameter
values used in this work are taken from Ref.~\cite{Wang:2024mgi}. For the two
vector-kaon configurations considered here, the calculated masses are
$m_{K^*(3\,{}^3S_1)}=1987$ MeV and $m_{K^*(2\,{}^3D_1)}=2121$ MeV.
These values exceed the BESIII central mass by 118 MeV and 252 MeV,
respectively, so neither pure state can be in accordance with the
observed resonance.

\begin{figure}[t]
\centering
\includegraphics[width=\columnwidth]{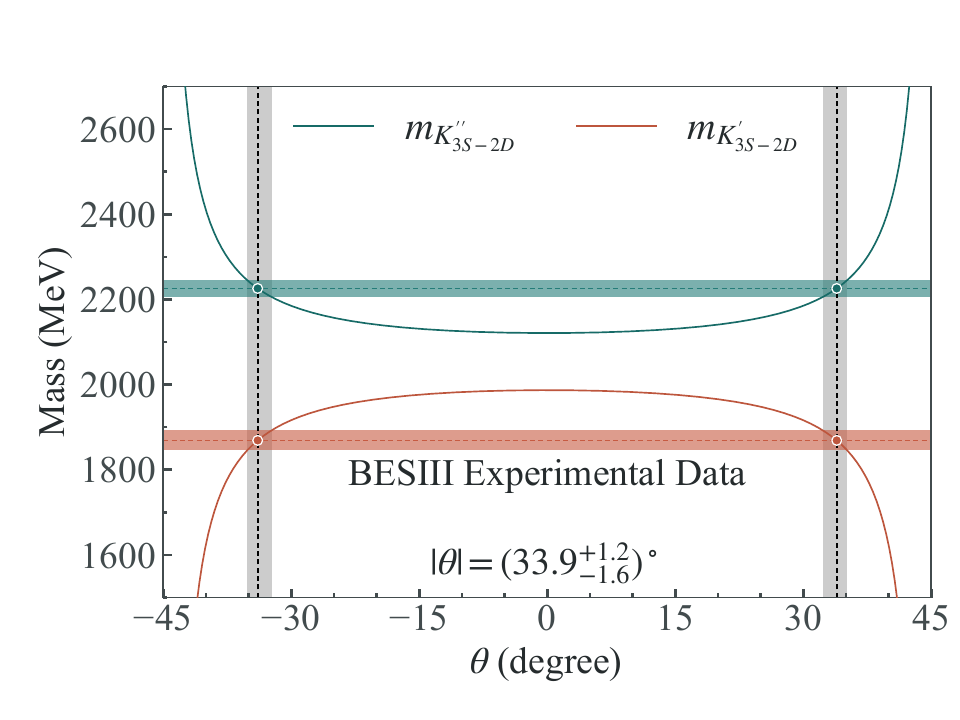}
\caption{Masses of the $K^{\prime}_{3S-2D}$ and
$K^{\prime\prime}_{3S-2D}$ states as functions of the mixing angle $\theta$. The red
band marks the BESIII mass interval \cite{BESIII:2026K1star1870}. The black bands indicate the allowed mixing-angle intervals for the positive and negative values of $\theta$, and the green band gives the mass interval of the partner state. The
curves use $m_{K(3\,{}^3S_1)}=1987$ MeV and
$m_{K(2\,{}^3D_1)}=2121$ MeV.}
\label{fig:sd_mass}
\end{figure}

Both configurations have $J^P=1^-$. We denote the two mixed states by
$K^{\prime}_{3S-2D}$ and $K^{\prime\prime}_{3S-2D}$. The former is
identified with the $K_1^*(1870)$, while the latter is its higher-mass
partner, denoted by $K_1^*(2225)$ in Sec.~\ref{section3}. The wave functions of the mixed states can be expressed as
\begingroup
\small
\setlength{\arraycolsep}{2pt}
\begin{equation}
\begin{pmatrix}
|K^{\prime}_{3S-2D}\rangle\\[3pt]
|K^{\prime\prime}_{3S-2D}\rangle
\end{pmatrix}
=
\begin{pmatrix}
\cos\theta & \sin\theta\\[3pt]
-\sin\theta & \cos\theta
\end{pmatrix}
\begin{pmatrix}
|K(3\,{}^3S_1)\rangle\\[3pt]
|K(2\,{}^3D_1)\rangle
\end{pmatrix}.
\label{eq:sd_states}
\end{equation}
\endgroup
Such $S$-$D$ mixing has been considered in several vector-meson flavors, including charmonium and bottomonium, and has been used to
interpret the $\omega$-like state $Y(2119)$
\cite{Rosner:2001mix,Wang:2019mix,Li:2021mix,Bai:2025sdmix}. Following
the mass-mixing prescription of Ref.~\cite{Bai:2025sdmix}, the
masses of the mixed states are
\begin{equation}
\begin{aligned}
m_{K^{\prime}_{3S-2D}}^2
={}&\frac{1}{2}\bigg[m_{K(3\,{}^3S_1)}^2
+m_{K(2\,{}^3D_1)}^2\\
&-\sqrt{\left(m_{K(2\,{}^3D_1)}^2-m_{K(3\,{}^3S_1)}^2\right)^2
\sec^2 2\theta}\bigg],\\
m_{K^{\prime\prime}_{3S-2D}}^2
={}&\frac{1}{2}\bigg[m_{K(3\,{}^3S_1)}^2
+m_{K(2\,{}^3D_1)}^2\\
&+\sqrt{\left(m_{K(2\,{}^3D_1)}^2-m_{K(3\,{}^3S_1)}^2\right)^2
\sec^2 2\theta}\bigg].
\end{aligned}
\label{eq:sd_masses}
\end{equation}
These expressions hold for \(|\theta| \lesssim 43^\circ\), which covers the
physical solutions considered below. Figure~\ref{fig:sd_mass} shows
their dependence on the mixing angle. We combine the statistical and
systematic uncertainties in quadrature and use
$m_{K^{\prime}_{3S-2D}}=1869.0\pm23.6$ MeV as input. The resulting
angle is
\begin{equation}
|\theta|=(33.9^{+1.2}_{-1.6})^\circ,
\label{eq:sd_angle}
\end{equation}
with the intervals $32.3^\circ\leq\theta\leq\textcolor{black}{35.2^\circ}$ and
$\textcolor{black}{-35.2^\circ}\leq\theta\leq-32.3^\circ$. The mass of the
$K^{\prime\prime}_{3S-2D}$ partner is
\begin{equation}
m_{K^{\prime\prime}_{3S-2D}}
=2225.7^{+19.6}_{-20.1}~\mathrm{MeV}.
\label{eq:sd_partner_mass}
\end{equation}

We next test this $3S$-$2D$ mixing assignment through its strong decays.
The OZI-allowed two-body widths are calculated with the QPC model \cite{Micu:1969qpc,LeYaouanc:1973qpc,Li:2022khh}.
In this model, a color-singlet quark-antiquark pair is created from
the vacuum with spin $S=1$ and orbital angular momentum $L=1$,
coupled to $J^{PC}=0^{++}$.
The relation of this phenomenological picture to microscopic decay
mechanisms has been studied in Ref.~\cite{Ackleh:1996qpc}.
Applications to strange mesons, higher charmonia, excited light vector
mesons, and high-spin kaons can be found in
Refs.~\cite{Barnes:2003strange,Blundell:1996strange,Barnes:2005charm,
Feng:2021qmt,Wang:2025lin}.

We evaluate the decay amplitudes using the MGI spatial wave functions
and adopt the pair-creation strengths $\gamma=10.5$ for $u\bar u$ and
$d\bar d$ and $\gamma_s=10.5/\sqrt{3}$ for $s\bar s$
\cite{Pang:2017dlw}. The helicity amplitudes
are projected onto partial waves $\mathcal M^{JL}$ using the
Jacob--Wick prescription \cite{Jacob:1959at}. In our normalization,
the two-body width is
\begin{equation}
\Gamma(A\to BC)=\frac{\pi P}{4m_A^2}
\sum_{JL}|\mathcal M^{JL}(P)|^2,
\label{eq:qpc_width}
\end{equation}
where the final-state momentum in the rest frame of $A$ is
\begin{equation}
P=\frac{\sqrt{\left[m_A^2-(m_B+m_C)^2\right]
\left[m_A^2-(m_B-m_C)^2\right]}}{2m_A}.
\label{eq:qpc_momentum}
\end{equation}
The total width is obtained by summing the included two-body channels.
For each mixed state, the two basis amplitudes are combined coherently
before taking their squared modulus \cite{Bai:2025sdmix}:
\begingroup
\small
\begin{equation}
\begin{aligned}
\Gamma_{K^{\prime}_{3S-2D}\to BC}
&=\frac{\pi}{4}\frac{|\bm P'|}{m_{K^{\prime}_{3S-2D}}^2}\sum_{JL}\\
&\quad\times\left|\cos\theta\,\mathcal M_{3\,{}^3S_1}^{JL}(\bm P')
+\sin\theta\,\mathcal M_{2\,{}^3D_1}^{JL}(\bm P')\right|^2,
\end{aligned}
\label{eq:sd_width_prime}
\end{equation}
\endgroup
\begingroup
\small
\begin{equation}
\begin{aligned}
\Gamma_{\textcolor{black}{K^{\prime\prime}_{3S-2D}}\to BC}
&=\frac{\pi}{4}\frac{|\bm P''|}{m_{\textcolor{black}{K^{\prime\prime}_{3S-2D}}}^2}\sum_{JL}\\
&\quad\times\left|-\sin\theta\,\mathcal M_{3\,{}^3S_1}^{JL}(\bm P'')
+\cos\theta\,\mathcal M_{2\,{}^3D_1}^{JL}(\bm P'')\right|^2.
\end{aligned}
\label{eq:sd_width_doubleprime}
\end{equation}
\endgroup
Here $\mathcal M_{3\,{}^3S_1}^{JL}$ and
$\mathcal M_{2\,{}^3D_1}^{JL}$ are the partial-wave amplitudes of the
two basis states. The indices $J$ and $L$ are the total spin and
relative orbital angular momentum of $B$ and $C$. Both amplitudes in
each equation are evaluated at the same physical parent mass. The
momentum magnitudes $|\bm P'|$ and $|\bm P''|$ follow from
Eq.~\eqref{eq:qpc_momentum} using the corresponding mixed-state mass.

For $K^{\prime}_{3S\text{-}2D}$, identified with $K_1^*(1870)$, the mass is fixed to the BESIII central value of 1869.0~MeV. At $\theta=+33.9^\circ$ and $\theta=-33.9^\circ$, the calculated total \textcolor{black}{widths} are 197.3~MeV and 198.1~MeV, respectively. These values are consistent with the BESIII result $167.3\pm34.0\pm24.2$~MeV within the experimental uncertainties~\cite{BESIII:2026K1star1870}. The mass fixes the mixing angle, while the width provides \textcolor{black}{independent support} for the $3S$-$2D$ mixing interpretation. We note that the width is obtained at the central values of the mass and mixing angle.

The partial widths of $K^{\prime}_{3S-2D}$ are listed in
Table~\ref{tab:sd_lower_decays}. Only channels with widths of at least 1 MeV in
either mixing-angle solution are listed. The $K\eta'$ channel, although it is the discovery mode, does
not reach this threshold, and its partial widths are 0.05 and
0.01 MeV for the positive- and negative-angle solutions, respectively.
With the total widths in Table~\ref{tab:sd_lower_decays}, the
corresponding branching fractions are $2\times10^{-4}$ and
$6\times10^{-5}$. For $\theta=+33.9^\circ$, the leading modes are
$K\pi$, $K_1(1650)\pi$, and $K(1460)\pi$, with widths of 33.8, 30.2,
and 24.2 MeV. For $\theta=-33.9^\circ$, the largest contributions
come from $K^*(1410)\pi$, $K_1(1650)\pi$, and $K^*(892)\pi$, with
widths of 56.7, 36.3, and 25.7 MeV. In particular, the $K\pi$ mode
is strongly suppressed in the negative-angle solution. Measurements of
these decay patterns can therefore help determine the sign of the
mixing angle in the convention of Eq.~\eqref{eq:sd_states}.

\begin{table}[!htbp]
\caption{Partial and total strong-decay widths of
$K^{\prime}_{3S-2D}$ in MeV, \textcolor{black}{for}
$\theta=+33.9^\circ$ and $-33.9^\circ$. The dash denotes a partial width
below 1 MeV.}
\label{tab:sd_lower_decays}
\centering
\footnotesize
\begin{ruledtabular}
\renewcommand{\arraystretch}{1.2}
\begin{tabular}{lrr}
Channel & $+33.9^\circ$ & $-33.9^\circ$ \\
\hline
$Ka_1(1260)$ & 8.7 & 6.9 \\
$Ka_2(1320)$ & -- & 1.4 \\
$Kb_1(1235)$ & 8.2 & 4.6 \\
$Kf_1(1285)$ & 2.9 & 5.0 \\
$Kf_2(1270)$ & -- & 1.7 \\
$Kh_1(1170)$ & 4.7 & 2.8 \\
$K_1(1270)\eta$ & 1.6 & 2.2 \\
$K\eta$ & 11.7 & -- \\
$K\eta(1295)$ & 8.3 & -- \\
$K_2^*(1430)\pi$ & 2.0 & 8.9 \\
$K_1(1650)\pi$ & 30.2 & 36.3 \\
$K(1630)\pi$ & 7.8 & 1.9 \\
$K(1460)\pi$ & 24.2 & -- \\
$K_1(1270)\pi$ & 3.5 & 4.9 \\
$K_1(1400)\pi$ & 10.8 & 2.1 \\
$K^*(892)\pi$ & 6.4 & 25.7 \\
$K^*(1410)\pi$ & 13.0 & 56.7 \\
$K\pi$ & 33.8 & -- \\
$K^*(892)\rho$ & 8.5 & 5.8 \\
$K\rho$ & 5.3 & 21.2 \\
$K^*(892)\omega$ & 3.0 & 2.0 \\
$K\omega$ & 1.7 & 6.8 \\
\hline
Total & 197.3 & 198.1 \\
\end{tabular}
\end{ruledtabular}
\end{table}

Table~\ref{tab:sd_higher_decays} gives the channels for the higher-mass $K^{\prime\prime}_{3S-2D}$, which has total widths of 874.8 MeV for
$\theta=+33.9^\circ$ and 826.7 MeV for $\theta=-33.9^\circ$ at the mass
of \textcolor{black}{2225.7} MeV. For the positive-angle solution, the largest
contributions are $K^*(892)\rho$, $K\rho$, and $K^*(892)\pi$, with
partial widths of 72.8, 63.4, and 62.2 MeV, respectively. For the
negative-angle solution, the leading modes are $K^*(892)\rho$,
$K^*(1410)\rho$, and $K_1(1400)\pi$, with widths of 106.0, 74.7, and
65.9 MeV. The partial-width patterns
provide specific channels for searches near 2.23 GeV and input for the
production analysis.

\begin{table}[!htbp]
\caption{Partial and total strong-decay widths of
$K^{\prime\prime}_{3S-2D}$ in MeV, \textcolor{black}{for}
$\theta=+33.9^\circ$ and $-33.9^\circ$. The dash denotes a partial width
below 1 MeV.}
\label{tab:sd_higher_decays}
\centering
\scriptsize
\setlength{\tabcolsep}{2pt}
\renewcommand{\arraystretch}{1.2}
\begin{ruledtabular}
\renewcommand{\arraystretch}{1.2}
\begin{tabular}{@{}lrr@{\hspace{8pt}}lrr@{}}
Channel & $+33.9^\circ$ & $-33.9^\circ$ &
Channel & $+33.9^\circ$ & $-33.9^\circ$ \\
\hline
$K^*(892)\phi$ & 1.5 & 1.4 & $K^*(892)\eta'$ & 3.8 & -- \\
$K^*(892)a_1(1260)$ & 34.3 & 45.0 & $K_2^*(1430)\pi$ & 27.3 & -- \\
$K^*(892)a_2(1320)$ & 4.4 & 4.8 & $K_2(1770)\pi$ & 59.6 & 62.0 \\
$K^*(892)b_1(1235)$ & 37.1 & 6.9 & $K_1(1270)\pi$ & 8.9 & 17.5 \\
$K^*(892)f_1(1285)$ & 2.6 & 9.9 & $K_1(1400)\pi$ & 49.5 & 65.9 \\
$K^*(892)f_2(1270)$ & 4.3 & 7.8 & $K^*(892)\pi$ & 62.2 & 1.7 \\
$K^*(892)h_1(1170)$ & 15.2 & 5.1 & $K^*(1410)\pi$ & 19.8 & -- \\
$K\phi$ & 4.6 & -- & $K\pi$ & 8.4 & 51.3 \\
$Ka_1(1260)$ & 17.7 & 25.6 & $K_3^*(1780)\pi$ & 3.8 & -- \\
$Ka_2(1320)$ & 29.5 & -- & $K_0^*(1430)\rho$ & 14.9 & 14.9 \\
$Kb_1(1235)$ & 43.4 & 55.4 & $K_2^*(1430)\rho$ & 4.9 & 5.6 \\
$Kf_1(1285)$ & 6.7 & 11.1 & $K_1(1270)\rho$ & 21.3 & 17.8 \\
$Kf_2'(1525)$ & 2.1 & -- & $K_1(1400)\rho$ & 7.9 & 35.7 \\
$Kf_2(1270)$ & 7.0 & -- & $K^*(892)\rho$ & 72.8 & 106.0 \\
$Kh_1(1380)$ & 2.8 & 2.7 & $K^*(1410)\rho$ & 36.0 & 74.7 \\
$Kh_1(1170)$ & 16.3 & 18.8 & $K\rho$ & 63.4 & 1.6 \\
$K\eta_2(1645)$ & 17.0 & 22.6 & $K\rho(1450)$ & 49.2 & -- \\
$K\pi_2(1670)$ & 20.5 & 27.2 & $K_0^*(1430)\omega$ & 4.4 & 4.4 \\
$K_2^*(1430)\eta$ & 10.8 & -- & $K_2^*(1430)\omega$ & 1.4 & 1.5 \\
$K_1(1400)\eta$ & 2.0 & 9.0 & $K_1(1270)\omega$ & 7.3 & 6.0 \\
$K^*(892)\eta$ & 2.3 & -- & $K_1(1400)\omega$ & 2.2 & 12.0 \\
$K^*(1410)\eta$ & 1.4 & -- & $K^*(892)\omega$ & 23.8 & 34.4 \\
$K\eta$ & 3.1 & 26.8 & $K^*(1410)\omega$ & 9.0 & 18.8 \\
$K^*(892)\eta(1295)$ & 2.0 & -- & $K\omega$ & 21.0 & -- \\
$K\eta(1295)$ & 2.4 & 9.0 &  &  &  \\
\hline
\multicolumn{6}{c}{Total: $\Gamma(\theta=+33.9^\circ)=874.8~\mathrm{MeV}$,
$\Gamma(\theta=-33.9^\circ)=826.7~\mathrm{MeV}$} \\
\end{tabular}
\end{ruledtabular}
\end{table}

\section{\texorpdfstring{PRODUCTION OF $K_1^*(1870)$ AND ITS
PARTNER STATE via $Kp$ reaction}{PRODUCTION OF THE K1*(1870) AND ITS PARTNER STATE}}
\label{section3}

For the production calculations in this section, we restrict the mixing angle to the positive-angle solution, $\theta=+33.9^\circ$, in the convention
of Eq.~\eqref{eq:sd_states}. This choice is motivated by the fact that this solution yields a larger $K_1^*(1870)\to K\eta'$ branching fraction, $2\times10^{-4}$, compared with the negative-angle value of $6\times10^{-5}$. All production couplings and cross sections quoted below
correspond to the positive-angle solution for both mixed states.

\begin{figure}[t]
\centering
\includegraphics[width=\columnwidth]{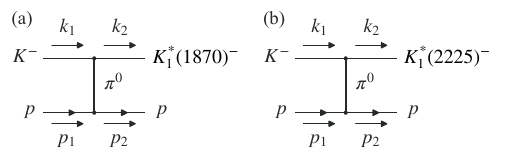}
\caption{Feynman diagrams for (a) $K^-p\to K_1^*(1870)^-p$ and (b)
$K^-p\to K_1^*(2225)^-p$ through $\pi^0$ exchange. The four-momenta of
the incoming kaon and proton are $k_1$ and $p_1$, and those of the
outgoing strange meson and proton are $k_2$ and $p_2$, respectively.}
\label{fig:production_diagrams}
\end{figure}

We study the production of the $K_1^*(1870)$ and its
$S$-$D$ mixed partner in the $K^-p$ reaction within the effective
Lagrangian approach \cite{Li:2026kprod}. It has been widely used in meson production calculations \cite{Oh:2006scalar,Ozaki:2009wp,Wang:2023lia,Wang:2025rvr}. For convenience, the higher-mass state $K^{\prime\prime}_{3S-2D}$ is
denoted by $K_1^*(2225)$ in this section. The Feynman diagrams
considered are shown in Fig.~\ref{fig:production_diagrams}. We retain only the dominant $t$-channel mechanism. The ground-state baryon poles in the $s$ channel lie far from the physical region, so their contributions are strongly suppressed. The $u$-channel contribution is also neglected because the relevant baryon-exchange couplings are expected to be small, as illustrated in Ref.~\cite{Li:2026kprod}. We therefore only calculate the corresponding $t$-channel amplitudes.

The possible exchanged mesons are selected from the kaon-containing
strong-decay channels calculated in Sec.~\ref{section2}. We examine
the contributions from the mesons allowed at these vertices. For $K_1^*(1870)$, the
pion exchange gives the dominant contribution, whereas the other
possible exchanged-meson contributions are negligible. Thus, only
the $\pi$-exchange channel is considered. For the $K_1^*(2225)$
partner, we likewise retain only pion exchange in the production
calculation.

We present the effective interaction Lagrangians associated
with the upper vertices in Fig.~\ref{fig:production_diagrams}. For the
vector state $K_1^*(1870)$, the $K_1^*(1870)K\pi$ interaction is
described by the conventional derivative coupling among one vector and
two pseudoscalar mesons. The corresponding Lagrangian reads
\begin{equation}
\begin{aligned}
\mathcal L_{K_1^*(1870)K\pi}
={}&g_{K_1^*(1870)K\pi}K_1^{*+;\alpha}(1870)
\left(K^-\partial_\alpha\pi^0-\pi^0\partial_\alpha K^-\right),
\end{aligned}
\label{eq:production_l1870kpi}
\end{equation}
where $g_{K_1^*(1870)K\pi}$ denotes the effective coupling constant of
the $K_1^*(1870)K\pi$ vertex. For the $K_1^*(2225)$ state, the
corresponding interaction is described by
\begin{equation}
\begin{aligned}
\mathcal L_{K_1^*(2225)K\pi}
={}&g_{K_1^*(2225)K\pi}K_1^{*+;\alpha}(2225)\left(K^-\partial_\alpha\pi^0-\pi^0\partial_\alpha K^-\right),
\end{aligned}
\label{eq:production_l2225kpi}
\end{equation}
where $g_{K_1^*(2225)K\pi}$ denotes the effective coupling constant of
the $K_1^*(2225)K\pi$ vertex. The coupling constants
$g_{K_1^*(1870)K\pi}$ and $g_{K_1^*(2225)K\pi}$ are determined from the
calculated strong-decay widths of $K_1^*(1870)\to K\pi$ and
$K_1^*(2225)\to K\pi$, respectively. The three-meson coupling constants
used in the production calculation are
$
g_{K_1^*(1870)K\pi}=1.07 $ and $
g_{K_1^*(2225)K\pi}=0.47$.

We next specify the pion-nucleon interaction at the lower vertices.
The phenomenological interaction is taken from the Nijmegen
extended-soft-core baryon-baryon model
\cite{Rijken:2010zzb,Nagels:2015lfa} and reads
\begin{equation}
\mathcal L_{\pi NN}
=\frac{f_{\pi NN}}{m_\pi}\bar N\gamma^\mu\gamma_5
\partial_\mu\widetilde\pi N.
\label{eq:production_lpion}
\end{equation}
Here, $N$ denotes the nucleon field and $\widetilde\pi=\bm\tau\cdot\bm\pi$,
while $m_\pi$ denotes the pion mass. The pion-nucleon coupling constant
$f_{\pi NN}=0.95$ is taken from Ref.~\cite{Rijken:2010zzb}.

Combining the upper three-meson vertices with the pion propagator and
the lower pion-nucleon vertex gives the tree-level pion-exchange
amplitudes for the two production reactions. We define the four-momentum
transfer as $q=p_2-p_1=k_1-k_2$ and the Mandelstam variable as $t=q^2$.
The exchanged momentum is directed from the upper vertex to the lower
vertex in Fig.~\ref{fig:production_diagrams}. For the $K_1^*(1870)$
production through pion exchange, the tree-level amplitude is written
as
\begin{equation}
\begin{aligned}
i\mathcal M_{K_1^*(1870)\pi K}
={}&-g_{K_1^*(1870)K\pi}\frac{f_{\pi NN}}{m_\pi}\bar u(p_2)\gamma^\mu\gamma_5u(p_1)q_\mu
\frac{i}{t-m_\pi^2}\\
&\times\epsilon^{*\alpha}(k_2)
(q_\alpha+k_{1\alpha})F_t(q).
\end{aligned}
\label{eq:production_amplitude_1870_pi}
\end{equation}
For the $K_1^*(2225)$ production through pion exchange, the
corresponding tree-level amplitude reads
\begin{equation}
\begin{aligned}
i\mathcal M_{K_1^*(2225)\pi K}
={}&-g_{K_1^*(2225)K\pi}\frac{f_{\pi NN}}{m_\pi}\bar u(p_2)\gamma^\mu\gamma_5u(p_1)q_\mu
\frac{i}{t-m_\pi^2}\\
&\times\epsilon^{*\alpha}(k_2)
(q_\alpha+k_{1\alpha})F_t(q).
\end{aligned}
\label{eq:production_amplitude_2225_pi}
\end{equation}
In Eqs.~\eqref{eq:production_amplitude_1870_pi} and
\eqref{eq:production_amplitude_2225_pi}, $\epsilon^\alpha(k_2)$ denotes
the polarization vector of the produced $K_1^*(1870)$ or
$K_1^*(2225)$, respectively, while $u(p_1)$ and $\bar u(p_2)$ are the
Dirac spinors of the incoming and outgoing nucleons. The
phenomenological function $F_t(q)$ accounts for the finite-size effects
at the interaction vertices. For $t$-channel meson exchange, we adopt
the form factor
\begin{equation}
F_t(q)=\left(\frac{\Lambda_t^2-m^2}{\Lambda_t^2-q^2}\right)^2,
\label{eq:production_form_factor}
\end{equation}
where $m$ represents the mass of the exchanged meson. We use
$\Lambda_t=1.5\pm0.2$ GeV, as determined in Ref.~\cite{Li:2026kprod} by
fitting the available $K^-p\to K_3^*(1780)p$ cross-section data. With
the same cutoff, that work also describes the available $K_2(1770)$ and
$K_2(1820)$ production data well. We therefore adopt this parameter
range for the present reactions.

At higher energies, we replace the $t$-channel pion Feynman
propagator with a Regge propagator \cite{Guidal:1997hy}. This
prescription accounts for the exchange of higher-spin and higher-mass
states along the pion trajectory. Following Ref.~\cite{Guidal:1997hy},
we omit the Regge phase factor, so that the replacement is
\begin{equation}
\frac{1}{t-m_\pi^2}\longrightarrow
\left(\frac{s}{s_{\mathrm{scale}}}\right)^{\alpha_\pi(t)}
\frac{\pi\alpha'_\pi}
{\Gamma[\alpha_\pi(t)+1]\sin[\pi\alpha_\pi(t)]}.
\label{eq:production_regge_pi}
\end{equation}
The pion Regge trajectory is
\begin{equation}
\alpha_\pi(t)=\alpha'_\pi(t-m_\pi^2),
\label{eq:production_trajectories}
\end{equation}
with slope $\alpha'_\pi=0.7~\mathrm{GeV}^{-2}$. We take
$s_{\mathrm{scale}}=1~\mathrm{GeV}^2$.

\begin{figure}[t]
\centering
\includegraphics[width=\columnwidth]{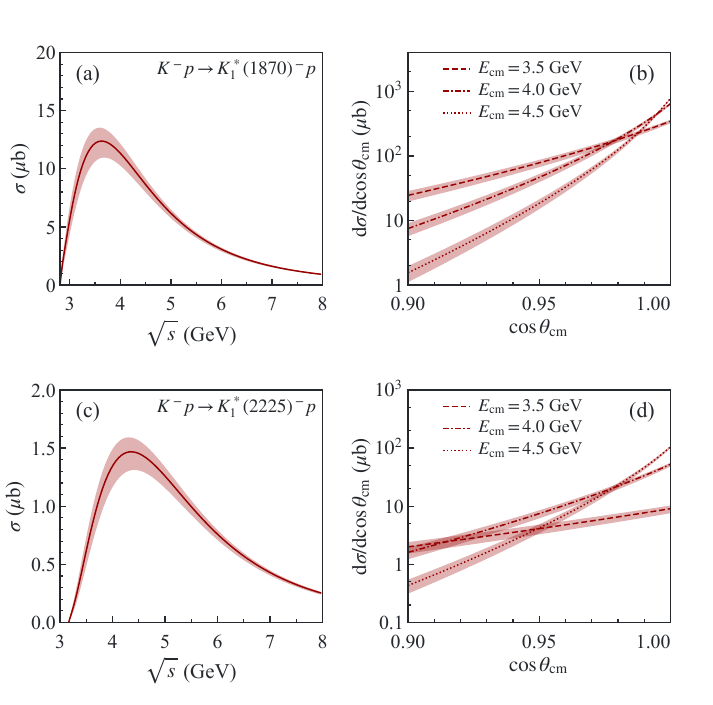}
\caption{Calculated production cross sections for
$K^-p\to K_1^*(1870)^-p$ in panels (a) and (b), and
$K^-p\to K_1^*(2225)^-p$ in panels (c) and (d).
The left panels show the total cross sections as functions of the
c.m. energy $\sqrt{s}$, while the right panels show the differential
cross sections in the forward-angle region at
$E_{\mathrm{cm}}=3.5$, 4.0, and 4.5 GeV, using dashed,
dash-dotted, and dotted curves, respectively. The red curves correspond
to $\Lambda_t=1.5$ GeV, and the shaded bands are obtained by varying
the cutoff within $\Lambda_t=1.5\pm0.2$ GeV
\cite{Li:2026kprod}.}
\label{fig:production_cross_sections}
\end{figure}

Within this framework, the differential cross section in the
center-of-mass (c.m.) frame is
\begin{equation}
\frac{d\sigma}{d\cos\theta_{\mathrm{cm}}}
=\frac{1}{32\pi s}\frac{|\bm k_2^{\mathrm{cm}}|}
{|\bm k_1^{\mathrm{cm}}|}
\left(\frac{1}{2}\sum_\lambda|\mathcal M|^2\right).
\label{eq:production_cross_section}
\end{equation}
Here, $s=(k_1+p_1)^2$, and $\theta_{\mathrm{cm}}$ denotes the angle
between the outgoing meson and the incoming $K^-$ beam in the c.m.
frame. The three-momenta $\bm k_1^{\mathrm{cm}}$ and
$\bm k_2^{\mathrm{cm}}$ refer to the initial kaon and final meson,
respectively. The sum runs over the spin and polarization states, and
the factor $1/2$ averages over the initial proton spin. The total cross
section follows by integrating over $-1\leq\cos\theta_{\mathrm{cm}}\leq1$.
The calculated total and differential cross sections are shown in
Fig.~\ref{fig:production_cross_sections}, with the $K_1^*(1870)$ results
in the upper row and the $K_1^*(2225)$ results in the lower row.

For the $K_1^*(1870)$ production, Fig.~\ref{fig:production_cross_sections}(a) shows a rapid
rise of the pion-exchange total cross section above the threshold
$\sqrt{s}\simeq2.8$ GeV. With $\Lambda_t=1.5$ GeV, the cross section
reaches a maximum of approximately $12.4~\mu\mathrm{b}$ at
$\sqrt{s}\simeq3.6$ GeV and then decreases with increasing energy. At
this energy, varying the cutoff from 1.3 to 1.7 GeV gives approximately
$11$--$14~\mu\mathrm{b}$. These bands describe the sensitivity to
the adopted cutoff range, rather than the full theoretical uncertainty.
The angular distributions in Fig.~\ref{fig:production_cross_sections}(b) are strongly
concentrated in the forward direction. At $\cos\theta_{\mathrm{cm}}=1$,
the central values of $d\sigma/d\cos\theta_{\mathrm{cm}}$ are
approximately 345, 644, and $772~\mu\mathrm{b}$ at
$E_{\mathrm{cm}}=3.5$, 4.0, and 4.5 GeV, respectively. The forward
values increase with energy while the total cross section decreases
above its maximum, which indicates that the angular distribution
becomes more sharply forward peaked at higher energies.

For the $K_1^*(2225)$ production, the pion-exchange total cross
section in Fig.~\ref{fig:production_cross_sections}(c) reaches approximately
$1.5~\mu\mathrm{b}$ at $\sqrt{s}\simeq4.4$ GeV for $\Lambda_t=1.5$
GeV and then decreases with increasing energy. At this energy, varying
$\Lambda_t$ from 1.3 to 1.7 GeV gives a cross section of approximately
$1.3$--$1.6~\mu\mathrm{b}$. The angular distributions in
Fig.~\ref{fig:production_cross_sections}(d) increase toward $\cos\theta_{\mathrm{cm}}=1$. At
this forward endpoint, the central differential cross sections are
approximately 9, 53, and $105~\mu\mathrm{b}$ at
$E_{\mathrm{cm}}=3.5$, 4.0, and 4.5 GeV, respectively.

\section{Summary}
\label{section4}

Motivated by the BESIII observation of the $K_1^*(1870)$ in the
$K\eta'$ system, we have investigated its interpretation as the
lower-mass state arising from $3\,{}^3S_1$--$2\,{}^3D_1$ mixing in the
kaon meson family. Using the MGI basis masses of 1987 and 2121 MeV, the
measured mass fixes the magnitude of the mixing angle to
$|\theta|=(33.9^{+1.2}_{-1.6})^\circ$ and gives an orthogonal partner
with mass $2225.7^{+19.6}_{-20.1}$ MeV. At the central mass and mixing
angle, the QPC model predicts total widths of 197.3 and 198.1 MeV for
the positive- and negative-angle solutions of the $K_1^*(1870)$,
respectively. Both are compatible with the BESIII width of
$167.3\pm34.0\pm24.2$ MeV. This agreement supports the mixed-state
assignment, while the present mass and total width do not determine the
sign of the mixing angle.

The predicted decay patterns provide additional tests. For the
positive-angle solution, the leading modes of the $K_1^*(1870)$ are
$K\pi$, $K_1(1650)\pi$, and $K(1460)\pi$, whereas $K^*(1410)\pi$,
$K_1(1650)\pi$, and $K^*(892)\pi$ dominate for the negative-angle
solution. The $K\eta'$ branching fractions are only
$\textcolor{black}{2\times10^{-4}}$ and $\textcolor{black}{6\times10^{-5}}$, respectively, despite the
observation of the state in this channel. This tension, already noted
in Sec.~\ref{section2}, calls for further study. The partner is
predicted to be broad, with total widths of 874.8 and 826.7 MeV for the
two solutions. Its substantial $K^*(892)\rho$ partial width in both
cases identifies this mode as a promising channel for testing the
partner prediction.

We have also calculated the pion-exchange production of the two mixed
states in $K^-p$ reactions within an effective Lagrangian approach with
a Reggeized propagator. We adopt the positive-angle solution, motivated
by its larger $K\eta'$ branching fraction. For $\Lambda_t=1.5$ GeV, the total
cross sections peak at approximately $12.4~\mu\mathrm{b}$ near
$\sqrt{s}=3.6$ GeV for the $K_1^*(1870)$ and at $1.5~\mu\mathrm{b}$
near $\sqrt{s}=4.4$ GeV for its partner. Both angular distributions
are forward peaked. The present study also demonstrates that the $S$-$D$
mixing scheme, widely applied in the charmonium and bottomonium
sectors, \textcolor{black}{can be applied to the $K_1^*(1870)$}, offering a
template for identifying similar mixed states in the kaon meson
family.

\begin{acknowledgments}

\textcolor{black}{T. Y. Li} would like to thank Cheng-Xi Liu for useful discussions. This work is supported by the Natural Science Foundation of Gansu Province (No. 26RCKA012, No. 25JRRA799), the National Natural Science Foundation of China under Grants No. 12335001 and No. 12247101, the ‘111 Center’ under Grant No. B20063, the Fundamental Research Funds for the Central Universities (No. lzujbky-2023-stlt01), and Lanzhou City High-Level Talent Funding. Z. Y. Bai is also supported by the Postdoctoral Fellowship Program
of CPSF under Grant Number GZB20260801. The authors used ChatGPT to assist with literature searches, numerical calculations, figure preparation, and language editing. The authors reviewed
the output and take responsibility for the content.

\end{acknowledgments}

\bibliographystyle{apsrev4-1}
\bibliography{ref}

\end{document}